\documentclass[12pt]{article}

\usepackage[mathscr]{eucal}
\usepackage{amsfonts}
\usepackage{amssymb}
\usepackage{amsthm}

\begin{document}

\begin{titlepage}
\vspace*{1.5cm}

\renewcommand{\thefootnote}{$\ddagger$}
\begin{center}

{\LARGE\bf  Massive twistor particle with spin }

\vspace{0.5cm}

{\LARGE\bf   generated by Souriau-Wess-Zumino term }

\vspace{0.5cm}

{\LARGE\bf   and its quantization}

\vspace{2cm}

{\large\bf Sergey Fedoruk},${}^{\dagger}$\footnote{\,\,On leave of absence from V.N.\,Karazin Kharkov National University, Ukraine}
\qquad
{\large\bf Jerzy Lukierski} ${}^{\ast}$

\vspace{1.5cm}

${}^{\dagger}${\it Bogoliubov  Laboratory of Theoretical Physics, JINR,}\\ {\it Joliot-Curie 6, 141980 Dubna, Moscow region, Russia} \\ \vspace{0.1cm}

{\tt fedoruk@theor.jinr.ru}\\ \vspace{0.5cm}

${}^{\ast}${\it Institute for Theoretical Physics, University of Wroc{\l}aw,}\\ {\it pl. Maxa Borna 9, 50-204 Wroc{\l}aw, Poland} \\ \vspace{0.1cm}

{\tt lukier@ift.uni.wroc.pl}\\

\vspace{2cm}

%04.10.2012

%\vspace{0.5cm}

\end{center} \vspace{0.2cm} %\vskip 0.6truecm \nopagebreak

\begin{abstract}
\noindent
We present new model of $D=4$ relativistic massive particle with spin and we describe its quantization.
The model is obtained by an extension of standard relativistic phase space description of massive spinless particle
by adding new topological Souriau-Wess-Zumino term which depends on spin fourvector variable.
We describe equivalently our model as given by the free two-twistor action with suitable
constraints. An important tool in our derivation is the spin-dependent twistor shift,
which modifies standard Penrose incidence relations.
The quantization of the model provides the wave function with correct mass and spin eigenvalues.

\end{abstract}



\newpage

\end{titlepage}

\setcounter{footnote}{0}

\setcounter{equation}{0}
\section{Introduction}

\quad\, In order to introduce in geometric way the spin degrees of freedom one has to enlarge the space-time description
of relativistic point particles. Well-known introduction of spin degrees of freedom is provided by superspace extension
of space-time, with anticommuting Grassmann algebra attached to each space-time point.
Other way of introducing the geometric spin degrees of freedom is to consider twistorial particle models,
with primary spinorial coordinates. The single twistor space has the degrees of freedom describing massless
particles with arbitrary helicity \cite{PenrM}--\cite{HugTod}.
%\cite{PenrM,Hugh,HugTod}.
In order to describe in twistor space
the massive particles with arbitrary spin one should consider particle models in two-twistor space
\cite{Hugh}--\cite{Bette}.
%\cite{Hugh,HugTod,Perj,Bette}.

The Penrose twistor approach \cite{Penr1,PenrM,Hugh,Penr2}
has been shown to be a powerful tool for the analysis of different point–like and extended objects.
In last time there is some renaissance of the twistor methods connected with successful application of twistors in description
of amplitudes in (super)Yang-Mills and (super)gravity theories (see, for example, \cite{Wit}--\cite{Mas}).
%\cite{Wit,Berkov,CSW,BCSW,Br,Ski,Mas}).
It should be added that the twistor approach has been considered mainly for massless (super)particles (see e.g. \cite{Sor} for approximately
complete list of more references on this subject),
but its application to massive particles, especially with non–zero spin, were investigated in rather limited number of papers
\cite{Perj,Bette}, \cite{SV}--\cite{MezRTown}.
%\cite{Perj,Bette,SV,FedZ2,BALM,Fed,AFLM,FFLM,MezRTown}).

The description of particles with a nonconformal mass parameter, and nonvanishing spin requires additional
degrees of freedom which has been studied in space-time as well as in the twistorial approach.
In space-time formalism one introduces
an additional Pauli-Lubanski spin fourvector $w_\mu$
which satisfies the subsidiary conditions \cite{Tod,Hugh,Bette}
\begin{equation}\label{sab-cond}
w_\mu p^\mu=0\,,\qquad w^2\equiv w_\mu w^\mu=-m^2j^2\,,
\end{equation}
with with relativistic spin-shell described by $j^2$ and fourmomenta satisfying the mass-shell condition $p^2=m^2$.
Alternatively, in twistor approach the two-twistor space is required to describe the phase space of massive particle with arbitrary spin,
and one construct from two twistors the composite spin fourvector $w_\mu$ satisfying the constraints (\ref{sab-cond}).

In our presentation we shall generalize from $D\,{=}\,3$ to $D\,{=}\,4$ the arguments of Mezincescu, Routh and Townsend  \cite{MezRTown},
who demonstrated that for $D=3$ massive particle the nonvanishing spin is generated in phase space $(x^\mu,p_\nu)$ by adding the term in the action described by the pullback
to the world-line of the following symplectic $D=3$ two-form
\begin{equation}\label{Om-3}
\Omega_2^{(D=3)}=\frac{s}{2(p^2)^{3/2}}\,\epsilon^{\mu\nu\rho}p_\mu dp_\nu\wedge dp_\rho\,,
\end{equation}
satisfying $d\Omega_2^{(D=3)}=$. It appears that such a term describes in $D\,{=}\,3$ action the Lorentz-Wess-Zumino (LWZ) term $\Omega_1^{(D=3)}$
which is the solution of the equation $\Omega_2^{(D=3)}=d\Omega_1^{(D=3)}$ \cite{Schon,MezTown}.
Calculating in two-twistor formulation the LWZ term one can see that it generates the twistor shift which modifies standard Penrose incidence relations
as follows
\begin{equation}\label{inc-3}
\omega_\alpha^i=x_{\alpha\beta}\lambda^{\beta i}+\frac{s}{m}\,\lambda_{\alpha}^{i}\,.
\end{equation}
Using modified incidence relations (\ref{inc-3})
one can obtain in the twistorial action of $D\,{=}\,3$ massive particle
with spin the kinetic term for twistors, which implies standard twistor Poisson brackets (PB).
Moreover as it was shown in \cite{MezRTown}, the eigenvalues of the Casimir operators of $D=3$ Poincare algebra corresponds
to massive states with the $D\,{=}\,3$ counterpart of spin $s$.
Note that twistorial shift in twistorial models is more important for massive particles,
because for $D\,{=}\,3,4$ massless particles it does not produce any change of the particle helicities \cite{SSTV}.

In our paper we provide an analogous scheme by introducing in place of (\ref{Om-3}) for  $D\,{=}\,4$ the symplectic two-form
introduced by Souriau \cite{Sour,Tod,Ku}
\begin{equation}\label{Sor-form-in}
\Omega_2^{(D=4)}=\frac{1}{2m^2}\,\epsilon_{\mu\nu\rho\sigma}w^\rho p^\sigma\left(
\frac{1}{m^2}\,dp^\mu \wedge dp^\nu + \frac{1}{w^2}\,dw^\mu \wedge dw^\nu\right)\,,
\end{equation}
where $w^\mu$ is the Pauli-Lubanski vector satisfying the relations (\ref{sab-cond}).
In Sect.\,2 we consider firstly the $D\,{=}\,4$ spinless massive particle and we recall that such a model can be formulated in three
equivalent ways (see e.g. \cite{FFLM})

-- by using relativistic phase space description ($x^\mu, p_\mu$)

-- by employing mixed space-time/spinor description (Shirafuji formulation \cite{Shir})

-- by using two-twistor framework.\\
We obtain that in $D=4$ two-twistor space our model is described by free action with added six constraints:
two related with mass-shell condition, three describing vanishing spin and sixth introducing
vanishing $U(1)$ charge.

In Sect.\,3 we add in $D=4$ space-time formulation the Souriau-Wess-Zumino (SWZ) topological term which
depends on the spin four-vector $w_\mu$ (see (\ref{sab-cond})). After passing to the spinorial description 
one can calculate the SWZ term by
the pullback to the world-line of the Souriau symplectic two-form (\ref{Sor-form-in}).
Subsequently, using spin dependent twistor shift we obtain the model depending on two-twistor coordinates and auxiliary spin
three-vector which span the coordinates of two-sphere. We shall recall how to derive from topological action such semi-dynamical spin variables,
which satisfy $SU(2)$ PB bracket relations. In two-twistor description the model is described by free bilinear action
with four first class and two second class constraints imposed by Lagrange multiplier method,
what leaves eight unconstrained physical degrees of freedom.
Further, in Sect.\,4, using the two-twistor formulation of our particle model,
we obtain the relativistic wave functions with mass and properly quantized spin values.
In final Sect.\,5 we summarize main results and point out some possible generalizations of presented scheme.

\setcounter{equation}{0}
\section{Massive spinless particle}

\quad\, The three equivalent descriptions of massive spinless particle are known
but we present them here in order to prepare the ground for the generalization in Sect.\,3 to the case of the massive particle with spin.

Relativistic phase space formulation of massive spinless particle is defined by well-known action
\begin{equation}\label{act-st-0}
\tilde S_1=\int \,d\tau
\Big[p_{\mu}\,\dot x^{\mu}+
e\left(p_{\mu}p^{\mu}-m^2 \right)\Big]\,.
\end{equation}
Here, $x^{\mu}(\tau)$, $\mu=0,1,2,3$ are the coordinates of position, $\dot x^{\mu}=dx^{\mu}/d\tau$ and
$p_{\mu}$ is fourvector of momenta. We use the metric with plus time signature, $\eta_{\mu\nu}={\rm diag}(+---)$.

In order to pass to mixed space-time/spinorial Shirafuji formulation we should use
the Cartan-Penrose formula expressing the relativistic fourmomenta
by a pair of Weyl commuting spinors ($k=1,2$)
\footnote{We shall use $D=4$ two-spinor notation, i.e.
$p_{\alpha\dot\beta}=\frac{1}{\sqrt{2}}\,\sigma^{\mu}_{\alpha\dot\beta}p_{\mu}$,
$p_{\mu}=\frac{1}{\sqrt{2}}\,\tilde\sigma_{\mu}^{\dot\beta\alpha}p_{\alpha\dot\beta}$, where
$(\sigma^\mu)_{\alpha\dot\alpha}=(1_2,\vec{\sigma})_{\alpha\dot\alpha}$,
$(\tilde\sigma^\mu)^{\dot\alpha\alpha}=\epsilon^{\alpha\beta}\epsilon^{\dot\alpha\dot\beta}(\sigma^\mu)_{\beta\dot\beta}=
(1_2,-\vec{\sigma})^{\dot\alpha\alpha}$, $\sigma^{\mu\nu}=\frac{1}{2}\,\sigma^{[\mu}\tilde\sigma^{\nu]}$, $\tilde\sigma^{\mu\nu}=\frac{1}{2}\,\tilde\sigma^{[\mu}\sigma^{\nu]}$
$\sigma^{\mu\nu}_{\alpha\beta}=\epsilon_{\beta\gamma}(\sigma^{\mu\nu})_{\alpha}{}^{\gamma}$,
$\tilde\sigma^{\mu\nu}_{\dot\alpha\dot\beta}=\epsilon_{\dot\beta\dot\gamma}(\tilde\sigma^{\mu\nu})^{\dot\gamma}{}_{\dot\alpha}$.
So, we have $p^{\dot\beta\alpha}p_{\alpha\dot\beta} = p^{\mu}p_{\mu}$. We
use weight coefficient in (anti)symmetrization, i.e. $A_{(\alpha}B_{\beta)}=\frac12\,(A_{\alpha}B_{\beta}+A_{\beta}B_{\alpha})$,
$A_{[\alpha}B_{\beta]}=\frac12\,(A_{\alpha}B_{\beta}-A_{\beta}B_{\alpha})$. }
\begin{equation}\label{P-spin-expr} 
p_{\alpha\dot\alpha} = \lambda^k_{\alpha}\bar\lambda_{\dot\alpha k}\,,
\end{equation}
where
\begin{equation}\label{P-spin-expr1}
\lambda^k_{\alpha}=\left(\lambda^1_{\alpha},\lambda^2_{\alpha} \right),\qquad
\bar\lambda_{\dot\alpha k}=(\overline{\lambda^k_{\alpha}})=\left(\bar\lambda_{\dot\alpha 1},\bar\lambda_{\dot\alpha 2} \right).
\end{equation}
Massive spinless particle dynamics is described by the extension of Shirafuji approach \cite{Shir}
\begin{equation}\label{act-shir-0}
\tilde S_2=\int \,d\tau
\Big[\lambda^k_{\alpha}\bar\lambda_{\dot\beta k}\,\dot x^{\dot\beta\alpha}+
g\left(\lambda^{\alpha k}\lambda_{\alpha k}-2M \right)+
\bar g\left(\bar\lambda^{\dot\alpha k}\bar\lambda_{\dot\alpha k}-2\bar M \right)
\Big]\,.
\end{equation}
where $x^{\dot\beta\alpha}=\frac{1}{\sqrt{2}}\,\tilde\sigma_{\mu}^{\dot\beta\alpha}x^{\mu}$
and $M$ is a complexified mass parameter.
In action (\ref{act-shir-0}) there are incorporated the mass-shell constraints
\footnote{We go up and down the indices $\alpha,\beta,\gamma,...$ and $i,j,k,...$ by antisymmetric tensors
$\epsilon_{\alpha\beta}$, $\epsilon_{ij}$, $\epsilon^{\alpha\beta}$, $\epsilon^{ij}$: $A_\alpha=\epsilon_{\alpha\beta}A^\beta$,
$A^\alpha=\epsilon^{\alpha\beta}A_\beta$, $A_i=\epsilon_{ij}A^j$,
$A^i=\epsilon^{ij}A_j$. We take $\epsilon_{12}=\epsilon^{21}=1$.
}
\begin{equation}\label{constr-lambda}
\lambda^{\alpha i}\lambda_{\alpha k}=M \,\delta^i_k \,,\qquad
\bar\lambda^{\dot\alpha i}\bar\lambda_{\dot\alpha k}=\bar M \,\delta^i_k
\end{equation}
or equivalently
\begin{equation}\label{constr-lambda1}
\lambda^{\alpha k}\lambda_{\beta k}=M \,\delta^\alpha_\beta \,,\qquad
\bar\lambda^{\dot\alpha k}\bar\lambda_{\dot\beta k}=\bar M \,\delta^{\dot\alpha}_{\dot\beta}\,.
\end{equation}
Due to the constrains (\ref{constr-lambda}) we have the following real mass-shell condition
($p^{\dot\beta\alpha}=\epsilon^{\alpha\gamma}\epsilon^{\dot\beta\dot\delta}p_{\gamma\dot\delta}$)
\begin{equation}\label{P-mass}
p^{\dot\beta\alpha}p_{\alpha\dot\beta} = 2\,|M|^2
\end{equation}
and comparing with (\ref{act-st-0}) we get
\begin{equation}\label{m-M}
m = \sqrt{2}\,|M|\,.
\end{equation}

The pair of spinors $\lambda^k_{\alpha}$, $\bar\lambda_{\dot\alpha k}$ describe one-half of two-twistor components.
Remaining  twistorial components are defined by the Penrose incidence relations (see e.g. \cite{Penr1,PenrM,Penr2})
\begin{equation}\label{mu-spin0}
\mu^{\dot\alpha k}=x^{\dot\alpha\beta}\lambda^k_{\beta}\,,\qquad
\bar\mu^{\alpha}_{k}=\bar\lambda_{\dot\beta k}x^{\dot\beta\alpha}\,.
\end{equation}
The relations (\ref{P-spin-expr}) and (\ref{mu-spin0}) link the Poisson brackets (PB) of
space-time and twistor space approaches.
Namely, when the relations (\ref{mu-spin0}) are satisfied  then
\begin{equation}\label{sim-str-0}
p_\mu\,\dot x^{\mu}=
\lambda^k_{\alpha}\bar\lambda_{\dot\beta k}\,\dot x^{\dot\beta\alpha}=
\lambda^k_{\alpha}\dot{\bar\mu}^{\alpha}_{k}+\bar\lambda_{\dot\alpha k}\dot\mu^{\dot\alpha k} + \mbox{(total derivative)}
\end{equation}
and we get the kinematic terms which lead to canonical PB in relativistic phase space as well as in two-twistor space.

If space-time coordinates are real twistor incident relations (\ref{mu-spin0}) lead to the following
conditions
\begin{equation}\label{cond-0}
\lambda^i_{\alpha}\bar\mu^{\alpha}_{k}-
\bar\lambda_{\dot\alpha k}\mu^{\dot\alpha i}=0\,,
\end{equation}
which should be imposed as the constraints. Thus, after using (\ref{sim-str-0}) 
the pure twistor formulation is described by the action (see also \cite{MezRTown})
\begin{equation}\label{act-tw-0}
\tilde S_3=\int \,d\tau
\Big[\lambda^k_{\alpha}\dot{\bar\mu}^{\alpha}_{k}+\bar\lambda_{\dot\alpha k}\dot\mu^{\dot\alpha k}+
gT+\bar g\bar T +\Lambda^r S^r +\Lambda S
\Big]
\end{equation}
incorporating the mass constraints (see also (\ref{constr-lambda}))
\begin{equation}\label{mass-constr}
T\equiv\lambda^{\alpha k}\lambda_{\alpha k}-2M\approx0\,,\qquad
\bar T\equiv\bar\lambda^{\dot\alpha k}\bar\lambda_{\dot\alpha k}-2\bar M\approx0
\end{equation}
and the $U(2)$ constraints
\begin{eqnarray}\label{Sr-0}
S^r&\equiv&-{\textstyle\frac{i}{2}}\left(\lambda^i_{\alpha}\bar\mu^{\alpha}_{k}-
\bar\lambda_{\dot\alpha k}\mu^{\dot\alpha i}\right)(\tau^r)_i{}^k\approx0\,,\qquad r=1,2,3\\[6pt]
S&\equiv&i\left(\lambda^i_{\alpha}\bar\mu^{\alpha}_{i}- \bar\lambda_{\dot\alpha i}\mu^{\dot\alpha i}\right)\approx0\,,\label{S-0}
\end{eqnarray}
which are the traceless and trace parts of the conditions (\ref{cond-0}) (in (\ref{Sr-0}) the $2\times 2$ matrices
$(\tau^r)_i{}^k$, $i,k=1,2$, $r=1,2,3$ are the usual Pauli matrices).

The action  (\ref{act-tw-0}) yields canonical twistor Poisson brackets
\begin{equation}\label{PB-tw-0}
\{ \bar\mu^{\alpha}_{k}, \lambda^j_{\beta} \}_{{}_P}=\delta^{\alpha}_{\beta}\delta_k^j\,,\qquad
\{ \mu^{\dot\alpha i} ,\bar\lambda_{\dot\beta k}\}_{{}_P}=\delta^{\dot\alpha}_{\dot\beta}\delta^k_j,.
\end{equation}
Then, nonvanishing Poisson brackets of the constraints (\ref{mass-constr}), (\ref{Sr-0}), (\ref{S-0}) are
\begin{equation}\label{PB-constr-0}
\{ S^p, S^r \}_{{}_P}=\epsilon^{prs}S^s\,,
\end{equation}
\begin{equation}\label{PB-constr-0a}
\{ S ,T\}_{{}_P}=2iT+4iM\,,\quad \{ S ,\bar T\}_{{}_P}=-2i\bar T-4i\bar M\,,
\end{equation}
where the constraints ($S^p$, $S$) describe $U(2)$ PB algebra. One can check easily
that we can choose four real constraints $S^r$, $(\bar MT+M\bar T)$ as first class constraints whereas
two real constraints $S$ and $i(\bar MT-M\bar T)$ are second class. We get therefore six unconstrained degrees of freedom
what coincides with number of degrees of freedom in standard space-time formulation (\ref{act-st-0})
of massive particle.

In twistor formulation the Poincare generators $p_\mu$ and $m_{\mu\nu}=x_\mu p_\nu-x_\nu p_\mu$ are represented by the expressions  (\ref{P-spin-expr})
and
\begin{equation}\label{M-tw}
m_{\mu\nu}=-\sigma_{\mu\nu}^{\alpha\beta}m_{\alpha\beta}+\tilde\sigma_{\mu\nu}^{\dot\alpha\dot\beta}\bar m_{\dot\alpha\dot\beta}\,,\qquad\qquad
m_{\alpha\beta}=\lambda^k_{(\alpha}\bar\mu_{\beta)k}\,,\qquad \bar m_{\dot\alpha\dot\beta}=\bar\lambda_{(\dot\alpha k}\mu_{\dot\beta)}^{ k}\,.
\end{equation}
Then, Pauli-Lubanski vector $w_\mu=\frac12\,\epsilon_{\mu\nu\lambda\rho}p^\nu m^{\lambda\rho}$ has the following twistor representation
\begin{equation}\label{W-tw}
w_{\alpha\dot\alpha}=S^r u^r_{\alpha\dot\alpha}\,,
\end{equation}
where $S^r$ are defined by (\ref{Sr-0}) and (see e.g. \cite{FedZ2})
\begin{equation}\label{u-expr}
u^r_{\alpha\dot\alpha} = \lambda^i_{\alpha}(\tau^r)_i{}^k\bar\lambda_{\dot\alpha k}\,.
\end{equation}
Due to equation (\ref{m-M}) and the constraints (\ref{mass-constr}) the vectors  (\ref{u-expr}) satisfy
\begin{equation}\label{u-norm}
u^r_{\mu}u^s{}^{\mu} = - m^2\delta^{rs}\,.
\end{equation}
Therefore due to the constraints (\ref{Sr-0}) and formulae (\ref{W-tw})-(\ref{u-norm}) in consistency with (\ref{sab-cond}) we have
\begin{equation}\label{W2-0}
p^{\mu}w_{\mu} =0\,,\qquad w^{\mu}w_{\mu} = - m^2S^{r}S^{r}\,,
\end{equation}
\begin{equation}\label{S2-0}
S^{r}S^{r}=j^2\,.
\end{equation}
In conclusion the spin of the massive particle described by the twistor action (\ref{act-tw-0}) vanishes,
i.e. we should put $j=0$.

\setcounter{equation}{0}
\section{Massive particle with spin and twistor shift}

\quad\, We define $D{=}4$ massive spin particle
in space-time formulation with help of the action
\begin{equation}\label{act-st-s}
S_1=\tilde S_1 + \int \Omega_1^{(D=4)}+ \int d\tau\left[l_1(p^{\mu}w_{\mu})+l_2(w^{\mu}w_{\mu} + m^2j^2) \right]\,,
\end{equation}
where first term $\tilde S_1$ is given by (\ref{act-st-0}), one-form $\Omega_1^{(D=4)}$ is defined
by Souriau symplectic two-form (\ref{Sor-form-in}) as follows
\begin{equation}\label{Sor-form}
\Omega_2^{(D=4)}=d\Omega_1^{(D=4)}\,,
\end{equation}
and the constraints on Pauli-Lubanski four-vector  $w^\mu$ are imposed by Lagrange multipliers.

Using the expressions (\ref{P-spin-expr}), (\ref{W-tw}) and the property that $M$, $\bar M$
are constants we obtain the following twistorial expression for Souriau two-form
\begin{equation}\label{Sor-tw}
\Omega_2^{(D=4)}=-{\textstyle\frac{i}{2M\bar M}}\,S^r (\tau^r)_i{}^k \Big(
\bar M\,d\lambda^{\alpha i} \wedge d\lambda_{\alpha k} + M\,d\bar\lambda^{\dot\alpha i} \wedge d\bar\lambda_{\dot\alpha k}\Big)\,,
\end{equation}
where the three-vector $S^r$ satisfies the constraint (\ref{S2-0}).

We recall here that in the theory of massive relativistic free fields with spin the Pauli-Lubanski four-vector satisfies the relations
(\ref{W2-0}) with $s^r$ described by the nondynamical matrix realization of $SU(2)$ algebra.
Further, in order to obtain that $\Omega_2^{(D=4)}$ in relation (\ref{Sor-tw}) satisfies the condition $d\Omega_2^{(D=4)}=0$ we shall postulate that
\begin{equation}\label{S-dot}
\dot S^r=0\qquad \rightarrow\qquad S^r=s^r\,,\quad s^rs^r=j^2
\end{equation}
with the variables $s^r\in \mathbb{S}^2$ as classical counterparts of quantum spin components endowed with
$SU(2)$ PB relation
\begin{equation}\label{S-br-s}
\{ s^p, s^r \}_{{}_P}=\epsilon^{prq}s^q\,.
\end{equation}
Using (\ref{S-dot}) one sees easily that Liouville one-form $\Omega_1^{(D=4)}$ satisfying (\ref{Sor-form})
takes the form
\begin{equation}\label{Lio-tw}
\Omega_1^{(D=4)}=-{\textstyle\frac{i}{2M\bar M}}\,s^r (\tau^r)_i{}^k \Big(
\bar M\,\lambda^{\alpha i} d\lambda_{\alpha k} + M\,\bar\lambda^{\dot\alpha i} d\bar\lambda_{\dot\alpha k}\Big)
\end{equation}
and the action (\ref{act-st-s}) becomes the following Shirafuji-like action
\begin{equation}\label{act-shir-s}
\begin{array}{rcl}
S_2&=&{\displaystyle \int \,d\tau
\Big[\lambda^k_{\alpha}\bar\lambda_{\dot\alpha k}\,\dot x^{\dot\alpha\alpha}+
g\left(\lambda^{\alpha k}\lambda_{\alpha k}-2M \right)+
\bar g\left(\bar\lambda^{\dot\alpha k}\bar\lambda_{\dot\alpha k}-2\bar M \right)
\Big]}\\[7pt]
&& {\textstyle -\frac{i}{2M\bar M}} {\displaystyle\int} d\tau\,
s^r (\tau^r)_i{}^k \Big(
\bar M\,\lambda^{\alpha i} \dot\lambda_{\alpha k} + M\,\bar\lambda^{\dot\alpha i} \dot{\bar\lambda}_{\dot\alpha k}\Big)\,.
\end{array}
\end{equation}
It appears that due to relation (\ref{W-tw})
the constraint $p^{\mu}w_{\mu}=0$ is valid as identity,
thus the action (\ref{act-shir-s}) becomes the sum of the action (\ref{act-shir-0}) and
the twistorial Souriau-Wess-Zumino topological term, represented by second integral in (\ref{act-shir-s}).

It should be stressed that the postulated PB relations (\ref{S-br-s}) can be derived from the dynamical
formulation if we supplement the action (\ref{act-shir-s}) with the following topological (Chern-Simons)
coupling term (see e.g. \cite{HoTo}--\cite{FedIvLech})
%\cite{HoTo,DuJaTr,FedIvLech})
\begin{equation}\label{de-act-shir-s}
\triangle \,S_2=\int \,d\tau
\Big[\mathcal{A}^r(s)\dot s^r+
l\left(s^r s^r-j^2 \right)
\Big]\,,
\end{equation}
where three-potential $\mathcal{A}^r(S)$ is such that
\begin{equation}\label{strengts}
{\mathcal F}^{rq}=\partial^r\mathcal{A}^q-\partial^q\mathcal{A}^r=-j\epsilon^{rqt}s^t/|s|^3\,.
\end{equation}
In order to derive the conditions $\dot s^r=0$ one should then pass to twistor formulation
and fix the local $SU(2)$ gauge which are generated by first class constraints defined below (see (\ref{Sr-s})).

Let us eliminate the space-time variables $x^\mu$ and pass to pure twistorial formulation
in two-twistor space. This requires to define
second twistorial spinors. As first attempt one can use the relations (\ref{mu-spin0})
as defining the second pair of Weyl twistors $\mu^{\dot\alpha k}$, but if we use the spinor variables $\lambda^k_{\beta}$, $\mu^{\dot\alpha k}$
the terms with derivatives in the action (\ref{act-shir-s})
\begin{equation}\label{sim-str-mixs}
\lambda^k_{\alpha}\bar\lambda_{\dot\beta k}\,\dot x^{\dot\beta\alpha}-{\textstyle\frac{i}{2M\bar M}} \,
s^r (\tau^r)_i{}^k \Big(
\bar M\,\lambda^{\alpha i} \dot\lambda_{\alpha k} + M\,\bar\lambda^{\dot\alpha i} \dot{\bar\lambda}_{\dot\alpha k}\Big)
\end{equation}
take the form
\begin{equation}\label{sim-str-s1}
\lambda^k_{\alpha}\dot{\bar\mu}^{\alpha}_{k}+\bar\lambda_{\dot\alpha k}\dot\mu^{\dot\alpha k}
-{\textstyle\frac{i}{2M\bar M}} \, s^r (\tau^r)_i{}^k \Big(
\bar M\,\lambda^{\alpha i} \dot\lambda_{\alpha k} + M\,\bar\lambda^{\dot\alpha i} \dot{\bar\lambda}_{\dot\alpha k}\Big)
+ \mbox{(total derivative)}\,.
\end{equation}
The kinetic terms given by (\ref{sim-str-s1}) show that the variables $\lambda^k_{\beta}$, $\mu^{\dot\alpha k}$ and
their complex conjugated do not satisfy the  canonical twistorial Poisson brackets.

In order to obtain the canonical twistorial PB we should redefine the half of twistor variables by
the following modified incidence relations
\begin{equation}\label{mu-spin-s}
\begin{array}{rcl}
\omega^{\dot\alpha k}&=&\mu^{\dot\alpha k}+{\textstyle\frac{i}{2\bar M}} \,s^r (\tau^r)_j{}^k \bar\lambda^{\dot\alpha j}=
x^{\dot\alpha\beta}\lambda^k_{\beta}+{\textstyle\frac{i}{2\bar M}} \,s^r (\tau^r)_j{}^k \bar\lambda^{\dot\alpha j}\,,\\[6pt]
\bar\omega^{\alpha}_{k}&=&\bar\mu^{\alpha}_{k}+{\textstyle\frac{i}{2M}} \,s^r (\tau^r)_k{}^j \lambda^{\alpha}_{j}=
\bar\lambda_{\dot\beta k}x^{\dot\beta\alpha}+{\textstyle\frac{i}{2M}} \,s^r (\tau^r)_k{}^j \lambda^{\alpha}_{j}\,.
\end{array}
\end{equation}
The formulae (\ref{mu-spin-s}) describe the spin-dependent twistor shift from Weyl spinors $\lambda^k_{\alpha}$, $\mu^{\dot\alpha k}$ to
$\lambda^k_{\alpha}$, $\omega^{\dot\alpha k}$.
It appears that subsequently the kinetic terms (\ref{sim-str-mixs})
take (even without (\ref{S-dot})) the standard form
\begin{equation}\label{sim-str-s}
\lambda^k_{\alpha}\dot{\bar\omega}^{\alpha}_{k}+\bar\lambda_{\dot\alpha k}\dot\omega^{\dot\alpha k}
+ \mbox{(total derivative)}\,.
\end{equation}
We see that the variables  ($\lambda^k_{\beta}$, $\omega^{\dot\alpha k}$) and ($\bar\lambda_{\dot\beta k}$, $\bar\omega^{\alpha}_{k}$)
are the canonical twistor variables for particle with spin and they are obtained by
the twistor shift applied to standard Penrose incidence relations for spinless particle
(compare (\ref{mu-spin-s}) with (\ref{mu-spin0})).

If the space-time coordinates are real, the twistor incidence relations (\ref{mu-spin-s})  lead to the following
conditions
\begin{equation}\label{cond-s}
\lambda^i_{\alpha}\bar\omega^{\alpha}_{k}-
\bar\lambda_{\dot\alpha k}\omega^{\dot\alpha i}=-is^r(\tau^r)_k{}^i\,,
\end{equation}
which generalize the constraints (\ref{Sr-0}) in the presence of nonvanishing spin variables $s^r$.
Thus, in two-twistor formulation we have the constraints (\ref{mass-constr})
and the modified constraints (\ref{Sr-0})-(\ref{S-0}) 
\footnote{
We denote by $V^r$, $V$ the expressions (\ref{Sr-0})-(\ref{S-0}) for $S^r$, $S$ with the replacement
of $\mu^{\dot\alpha k}$ by $\omega^{\dot\alpha k}$ (see (\ref{mu-spin-s})). The constraints (\ref{Sr-0})
are additionally modified by inhomogeneous terms proportional to $s^r$.
}
\begin{eqnarray}\label{Sr-s}
\mathcal{V}^r\equiv V^r+s^r&\equiv&-{\textstyle\frac{i}{2}}\left(\lambda^i_{\alpha}\bar\omega^{\alpha}_{k}-
\bar\lambda_{\dot\alpha k}\omega^{\dot\alpha i}\right)(\tau^r)_i{}^k+s^r\approx0\,,\qquad r=1,2,3\\[6pt]
V&\equiv&i\left(\lambda^i_{\alpha}\bar\omega^{\alpha}_{i}- \bar\lambda_{\dot\alpha i}\omega^{\dot\alpha i}\right)\approx0\,,\label{S-s}
\end{eqnarray}
which traceless and trace parts of the conditions (\ref{cond-s}) supplemented by the condition (\ref{S2-0}).
Thus, pure twistor formulation is described by the action
\begin{equation}\label{act-tw-s}
S_3=\int \,d\tau
\Big[\lambda^k_{\alpha}\dot{\bar\omega}^{\alpha}_{k}+\bar\lambda_{\dot\alpha k}\dot\omega^{\dot\alpha k}+
gT+\bar g\bar T +\Lambda^r (V^r+s^r) +\Lambda V
\Big]\,.
\end{equation}
Semi-dynamical variables $s^r$ satisfy
the PB (\ref{S-br-s}) and can be described by the action (\ref{de-act-shir-s});  all the constraints in the model are introduced
by using Lagrange multipliers.

In the formulation (\ref{act-tw-s}) of our model Poincare generators are given by the expressions  (\ref{P-spin-expr})
and Lorentz generators are
\begin{equation}\label{M-tw-s}
M_{\alpha\beta}=\lambda^k_{(\alpha}\bar\omega_{\beta)k}\,,\qquad \bar M_{\dot\alpha\dot\beta}=\bar\lambda_{(\dot\alpha k}\omega_{\dot\beta)}^{ k}\,.
\end{equation}
The Pauli-Lubanski vector $W_{\alpha\dot\alpha}=ip_{\alpha}^{\dot\beta} \bar M_{\dot\alpha\dot\beta}-
ip_{\dot\alpha}^{\beta} M_{\alpha\beta}$ has the following twistor representation
\begin{equation}\label{W-tw-s}
W_{\alpha\dot\alpha}=V^r u^r_{\alpha\dot\alpha}\,,
\end{equation}
where $V^r$ are defined in (\ref{Sr-s}) and $u^r_{\alpha\dot\alpha}$ by (\ref{u-expr}).
Further due to the constraints (\ref{Sr-s}) and relation (\ref{S2-0})
we get
\begin{equation}\label{W2-s}
W^{\mu}W_{\mu} = - m^2(V^{r}V^{r})=- m^2(s^{r}s^{r})=- m^2s^{2}\,.
\end{equation}

The action  (\ref{act-tw-s}) yields the canonical twistor Poisson brackets
\begin{equation}\label{PB-tw-s}
\{ \bar\omega^{\alpha}_{k}, \lambda^j_{\beta} \}_{{}_P}=\delta^{\alpha}_{\beta}\delta_k^j\,,\qquad
\{ \omega^{\dot\alpha i} ,\bar\lambda_{\dot\beta k}\}_{{}_P}=\delta^{\dot\alpha}_{\dot\beta}\delta^k_j\,.
\end{equation}
The twistorial PB  of the quantities $V^r$ are the same as these for $s^r$ in (\ref{S-br-s})
\begin{equation}\label{PB-V-s}
\{ V^p, V^r \}_{{}_P}=\epsilon^{prq}V^q
\end{equation}
what will provide the relations (\ref{Sr-s}) as first class constraints.
Because twistor coordinates and variables $s^r$ are kinematically independent,
nonvanishing Poisson brackets of all constraints (\ref{mass-constr}), (\ref{Sr-s}), (\ref{S-s}) are the following
\begin{equation}\label{PB-cV-s}
\{ \mathcal{V}^p,\mathcal{V}^r \}_{{}_P}=\epsilon^{prq}\mathcal{V}^q\,,
\end{equation}
\begin{equation}\label{PB-constr-s}
\{ V ,T\}_{{}_P}=2iT+4iM\,,\quad \{ V ,\bar T\}_{{}_P}=-2i\bar T-4i\bar M\,.
\end{equation}
We see that in present model four constraints are first class: three constraints $\mathcal{V}^r$
and the constraint $(\bar MT+M\bar T)$.
Other two constraints $V$ and $i(\bar MT-M\bar T)$ are second class.
In comparison with spinless case, we have additional two degrees of freedom in $s^r$,
describing spin degrees of freedom and the number of unconstrained degrees is $18-10=8$.

\setcounter{equation}{0}
\section{Quantization and field twistor transform}

We obtained the system, which is described in phase space by the variables $\lambda^j_{\alpha}$, $\bar\lambda_{\dot\alpha k}$, $\bar\omega^{\alpha}_{k}$,
$\omega^{\dot\alpha i}$, $s^r$, with canonical brackets (\ref{PB-tw-s}), (\ref{S-br-s}) and the constraints
$T$, $\bar T$ (see (\ref{mass-constr})), $\mathcal{V}^r$ (see (\ref{Sr-s})) and $V$ (see (\ref{S-s})).
The constraints $V$ and $i(\bar MT-M\bar T)$ are second class.
We shall introduce the gauge fixing condition
\begin{equation}\label{gauge}
G=\lambda^i_{\alpha}\bar\omega^{\alpha}_{i}+ \bar\lambda_{\dot\alpha i}\omega^{\dot\alpha i}\approx0
\end{equation}
for the local gauge transformations generated by the constraint $\bar MT+M\bar T$, i.e. we get second pair of second class constraints.
After introducing Dirac bracket for the second class constraints
\Big($V$, $i(\bar MT-M\bar T)$\Big), \Big($\bar MT+M\bar T$, $G$\Big) will should only impose three first class constraints $\mathcal{V}^r$.

Nonvanishing PBs of the constraint  (\ref{gauge}) are
\begin{equation}\label{PB-G-s1}
\{ G ,T\}_{{}_P}=2T+4M\,,\quad \{ G ,\bar T\}_{{}_P}=2\bar T+4\bar M\,.
\end{equation}
Then, the Dirac brackets (DB) for second class constraints $V$, $G$ and
\begin{equation}\label{F-s1}
F_1= \bar MT+M\bar T\,,\quad F_2=i(\bar MT-M\bar T)
\end{equation}
are given by formula
\begin{equation}\label{DB}
\!\!\!\!\!\!\begin{array}{c}
\{ A,B \}_{{}_D}=\{ A,B \}_{{}_P}+ \\[5pt]
{\textstyle\frac{1}{8M\bar M}}\Big[
\{ A,G \}_{{}_P}\{ F_1,B \}_{{}_P}-\{ A,F_1 \}_{{}_P}\{ G,B \}_{{}_P}
-\{ A,V \}_{{}_P}\{ F_2,B \}_{{}_P}+\{ A,F_2 \}_{{}_P}\{ V,B \}_{{}_P}
\Big]\,.
\end{array}
\end{equation}
The DBs for twistor spinor components take the form
\begin{equation}\label{DB-tw-l}
\{ \lambda^k_{\alpha}, \lambda^j_{\beta} \}_{{}_D}=
\{ \bar\lambda_{\dot\alpha k} ,\bar\lambda_{\dot\beta j}\}_{{}_D}=
\{ \lambda^k_{\alpha} ,\bar\lambda_{\dot\beta j}\}_{{}_D}=0\,,
\end{equation}
\begin{equation}\label{DB-tw-2}
\{ \bar\omega^{\alpha}_{k}, \lambda^j_{\beta} \}_{{}_D}=\delta^{\alpha}_{\beta}\delta_k^j+
{\textstyle\frac{1}{2M}}\lambda^{\alpha}_{k}\lambda^j_{\beta}\,,\qquad
\{ \omega^{\dot\alpha k} ,\bar\lambda_{\dot\beta j}\}_{{}_D}=\delta^{\dot\alpha}_{\dot\beta}\delta^k_j-
{\textstyle\frac{1}{2\bar M}}\bar\lambda^{\dot\alpha k} \bar\lambda_{\dot\beta j}\,,
\end{equation}
\begin{equation}\label{DB-tw-3}
\{  \omega^{\dot\alpha k}, \lambda^j_{\beta} \}_{{}_D}=0\,,\qquad
\{ \bar\omega^{\alpha}_{k},\bar\lambda_{\dot\beta j}\}_{{}_D}=0\,,
\end{equation}
\begin{equation}\label{DB-tw-4}
\{ \bar\omega^{\alpha}_{k}, \bar\omega^{\beta}_{j} \}_{{}_D}=-{\textstyle\frac{1}{M}}
\left(\lambda^{\alpha}_{k}\bar\omega^{\beta}_{j}-\lambda_j^{\beta} \bar\omega^{\alpha}_{k}\right)\,,\qquad
\{ \omega^{\dot\alpha k} ,\omega^{\dot\beta j}\}_{{}_D}=
{\textstyle\frac{1}{\bar M}}\left(\bar\lambda^{\dot\alpha k}\omega^{\dot\beta j}- \bar\lambda^{\dot\beta j}\omega^{\dot\alpha k}\right),
\end{equation}
\begin{equation}\label{DB-tw-5}
\{ \bar\omega^{\alpha}_{k}, \omega^{\dot\beta j}\}_{{}_D}=0\,.
\end{equation}

Further we consider $(\lambda,\bar\lambda)$-coordinate representation. In such spinorial Schr\"{o}dinger representation for the commutator
algebra obtained by quantization of DB (\ref{DB-tw-l})-(\ref{DB-tw-5})  the spinorial momentum operators
under suitable ordering ($\lambda$'s on the left, $\omega$'s on the right)
are realized in the following way
\begin{equation}\label{op-om}
\hat{\bar\omega}{}^{\alpha}_{k}=i\frac{\partial}{\partial\lambda_{\alpha}^{k}}+
\frac{i}{2M}\,\lambda_{\alpha}^{k}\,\lambda_{\beta}^{j} \frac{\partial}{\partial\lambda_{\beta}^{j}}\,,\qquad
\hat{\omega}{}^{\dot\alpha k}=i\frac{\partial}{\partial\bar\lambda_{\dot\alpha k}}-
\frac{i}{2\bar M}\,\bar\lambda_{\dot\alpha k}\,\bar\lambda_{\dot\beta j} \frac{\partial}{\partial\bar\lambda_{\dot\beta j}}\,.
\end{equation}
It is important that second terms in the operators (\ref{op-om}) do not contribute to the realization of quantum counterpart
$\hat V^r$ of the quantities $V^r$ (see (\ref{Sr-s})):
\begin{equation}\label{S-qu}
D^r\equiv \hat V^r={\frac{1}{2}}\left(\lambda^i_{\alpha}\frac{\partial}{\partial\lambda_{\alpha}^{k}}-
\bar\lambda_{\dot\alpha k}\frac{\partial}{\partial\bar\lambda_{\dot\alpha i}}\right)(\tau^r)_i{}^k\,.
\end{equation}
After quantization $s^r \to \hat s^r$ of the classical PB algebra (\ref{S-br-s}) we get the $SU(2)$ algebra
\begin{equation}\label{S-br-qu}
[ \hat s^p, \hat s^r ]=i\epsilon^{prq}\hat s^q\,,
\end{equation}
with classical constraint (\ref{S2-0}) becoming an operator identity
\begin{equation}\label{ss-s2-qu-cl}
\hat s^r\hat s^r=j^2\,.
\end{equation}
Because the quantum constraint (\ref{ss-s2-qu-cl}) describe the
eigenvalue condition of $SU(2)$ Casimir operator, for the unitary finite-dimensional representations of spin algebra (\ref{S-br-qu})
the value of $j^2$ are quantized in known way
\begin{equation}\label{ss-s2-qu}
j^2=J(J+1)\,,
\end{equation}
where $J$ is a non-negative half-integer number, i.e. $2J\in \mathbb{N}$.

{}For fixed $J$ the operators $\hat s^r$ are realized as $(2J+1)\times (2J+1)$ matrices.\footnote{
The constraints (\ref{Sr-s}) were already proposed in \cite{FedZ2}, however with the Schwinger realization
of the algebra (\ref{S-br-qu}) in terms of supplementary oscillators.}
Therefore,
twistor wave function of massive particle of spin $J$ has $(2J+1)$ components which are functions of $\lambda^i_{\alpha}$, $\bar\lambda_{\dot\alpha i}$,
constrained by strong conditions (\ref{constr-lambda}).
Because $\hat s^r\hat s^r$ commutes with $\hat s^3$, the wave function for fixed spin $J$ still depends on eigenvalues
$\mathcal{J}=(-J,-J+1,\ldots,J-1,J)$ of the spin projection $\hat s^3$. The wave function
($\lambda\equiv \lambda_\alpha^i$, $\bar\lambda=\bar\lambda_{\dot \alpha i}$)
\begin{equation}\label{wf-qu}
\Psi^{(J)}_{\mathcal{J}}=\Psi^{(J)}_{\mathcal{J}}(\lambda,\bar\lambda)\,,
\end{equation}
satisfies the matrix equations
\begin{equation}\label{Sr-qu}
(D^r+\hat S^r)\Psi^{(J)}=0\,,
\end{equation}
which is the quantum counterpart of the first class constraints (\ref{Sr-s}). We get the equations
\begin{equation}\label{Sr-qu-v}
D^rD^r\,\Psi^{(J)}_{\mathcal{J}}=J(J+1)\,\Psi^{(J)}_{\mathcal{J}}\,,\qquad D^3\,\Psi^{(J)}_{\mathcal{J}}=-\mathcal{J}\,\Psi^{(J)}_{\mathcal{J}}\,.
\end{equation}

{}From (\ref{S-qu}) follows that $D^r$ are the $SU(2)$ generators acting on indices
$i,k$ of twistor spinors $\lambda^i_{\alpha}$, $\bar\lambda_{\dot\alpha k}$
and $\hat s^r$  are the $SU(2)$ $(2J+1)\times (2J+1)$ matrix representation 
acting on index $\mathcal{J}$ of twistor wave function $\Psi^{(J)}_{\mathcal{J}}$. The formula
(\ref{Sr-qu}) links the parameters of both transformations and provide the following
transformations of the twistor wave function under $SU(2)$ local transformations
($\lambda^\prime{}^i_{\alpha}=h^i_k\lambda^k_{\alpha}$; $h\in SU(2)$):
\begin{equation}\label{su2-tr}
\Psi^{\prime(J)}_{\,\,\mathcal{J}}(\lambda^\prime)=\mathbf{D}^{(J)}_{\mathcal{J}\mathcal{K}}(h)\,\Psi^{(J)}_{\mathcal{K}}(\lambda)\,,
\end{equation}
where $\mathbf{D}^{(J)}_{\mathcal{J}\mathcal{K}}(h)$
is the matrix of irreducible $SU(2)$ representation of weight $J$.
We can represent equivalently the index ${\mathcal{J}}=-J,-J+1,\ldots,J$ as obtained by symmetrized $2J$ two-component spinor indices $i,j,k,...$
describing fundamental representation of the $SU(2)$ algebra  (\ref{S-br-qu})
and we get the twistor wave function as symmetric multispinor wave function
$\Psi^{(J)}_{\mathcal{J}}(\lambda)=\Psi^{(J)}_{(i_1...i_{2J})}(\lambda)$.

The space-time fields are obtained from twistor fields (\ref{wf-qu}) by integral transform containing
massive generalization of field twistor transform \cite{PenrM,Penr2,GaHoTo,FedZ2,Fed}. Such transform
is obtained if we construct $SU(2)$ invariant quantities by contraction of
the twistor fields (\ref{wf-qu}) with symmetrized multispinor indices $(i_1...i_{2J})$ with $\lambda^i_{\alpha}$, $\bar\lambda_{\dot\alpha i}$ and performing integral
with $SU(2)$-invariant measure with build-in mass-shell condition
\begin{equation}\label{meas}
d\mu_6(\lambda,\bar\lambda)=d^4 \lambda d^4\, \bar\lambda\, \delta(\lambda^{\alpha k}\lambda_{\alpha k}-2M)\,
\delta(\bar\lambda^{\dot\alpha k}\bar\lambda_{\dot\alpha k}-2\bar M)
\end{equation}
We use the Fourier transform with exponent $e^{ix^\mu p_\mu}$ containing the four-momentum which is expressed by bilinear twistor formula
(\ref{P-spin-expr}). The twistorial field with $2J$ $SU(2)$ indices produces by the suitable integration
with measure (\ref{meas}) the collection of $2J{+}1$ multispinor space-time fields
with Lorentz multispinor indices
\begin{equation}\label{st-fields}
\begin{array}{rcl}
\Phi^{(2J,0)}_{\alpha_1\ldots\alpha_{2J}}(x)&=& {\displaystyle\int d\mu_6(\lambda,\bar\lambda)\, e^{ix^{\dot\gamma\gamma} \lambda^k_{\gamma}\bar\lambda_{\dot\gamma k}}
\lambda^{i_1}_{\alpha_1}\ldots\lambda^{i_{2J}}_{\alpha_{2J}}\Psi^{(J)}_{i_1\ldots i_{2J}}(\lambda,\bar\lambda)}\,,\\[6pt]
\Phi^{(2J-1,1)}_{\alpha_1\ldots\alpha_{2J-1}}{}^{\dot\beta_1}(x)&=&{\displaystyle
\int d\mu_6(\lambda,\bar\lambda) \, e^{ix^{\dot\gamma\gamma} \lambda^k_{\gamma}\bar\lambda_{\dot\gamma k}}
\lambda^{i_1}_{\alpha_1}\ldots\lambda^{i_{2J}}_{\alpha_{2J-1}}\bar\lambda^{\dot\beta_1 i_{2J}}\Psi^{(J)}_{i_1\ldots i_{2J}}(\lambda,\bar\lambda)}\,,\\[6pt]
&& \hspace{2cm}..............\\[6pt]
\Phi^{(0,2J)}{}^{\dot\beta_1\ldots\dot\beta_{2J-1}}(x)&=&{\displaystyle
\int d\mu_6(\lambda,\bar\lambda) \, e^{ix^{\dot\gamma\gamma} \lambda^k_{\gamma}\bar\lambda_{\dot\gamma k}}
\bar\lambda^{\dot\beta_1 i_{1}}\ldots\bar\lambda^{\dot\beta_{2J} i_{2J}}\Psi^{(J)}_{i_1\ldots i_{2J}}(\lambda,\bar\lambda)}\,.
\end{array}
\end{equation}
In general case the wave functions (\ref{st-fields}) contain $n$ undotted symmetrized indices and $(2J{-}n)$ dotted symmetrized ones ($n=0,1,\ldots,2J$).
These space-time fields satisfy massive Dirac-like equations which reproduce in two-spinor notations the Bargmann-Wigner
fields.

Let us illustrate below the cases with lowest spins $J=0,\frac12,1$.

{\bf Spin 0:}
In this case twistor wave function $\Psi(\lambda,\bar\lambda)$ is a scalar field. Integral transform (\ref{st-fields}) gives us the scalar space-time field
\begin{equation}\label{sc-tr}
\Phi^{(0,0)}(x)= {\displaystyle\int d\mu_6(\lambda,\bar\lambda) \, e^{ix^{\dot\gamma\gamma} \lambda^k_{\gamma}\bar\lambda_{\dot\gamma k}}
\Psi^{(0)}(\lambda,\bar\lambda)}\,,
\end{equation}
which due to (\ref{constr-lambda})-(\ref{m-M}) satisfies the Klein-Gordon equation
\begin{equation}\label{KG-eq}
(\partial^\mu\partial_\mu+m^2)\,\Phi^{(0,0)}(x)= 0\,,
\end{equation}
i.e. describes in space-time the relativistic particle with mass $m$ and spin zero.

{\bf Spin 1/2:}
In this case due to integral transformations (\ref{st-fields}) we obtain two Weyl spinor fields
\begin{equation}\label{Ws12-tr}
\begin{array}{rcl}
\Phi^{(1,0)}_{\alpha}(x)&=& {\displaystyle\int d\mu_6(\lambda,\bar\lambda)\,  e^{ix^{\dot\gamma\gamma} \lambda^k_{\gamma}\bar\lambda_{\dot\gamma k}}
\lambda^{i}_{\alpha}\Psi^{(1/2)}_{i}(\lambda,\bar\lambda)}\,,\\[6pt]
\Phi^{(0,1)}{}^{\dot\beta}(x)&=&{\displaystyle
\int d\mu_6(\lambda,\bar\lambda)\, e^{ix^{\dot\gamma\gamma} \lambda^k_{\gamma}\bar\lambda_{\dot\gamma k}}
\bar\lambda^{\dot\beta i}\Psi^{(1/2)}_{i}(\lambda,\bar\lambda)}\,.
\end{array}
\end{equation}
These space-time fields due to algebraic properties of Weyl spinors satisfy the following generalized Dirac equations with complex mass $M$
\begin{equation}\label{Dir-eq}
i\partial^{\dot\beta\alpha}\Phi^{(1,0)}_{\alpha}+M\Phi^{(0,1)}{}^{\dot\beta}= 0\,,\qquad
i\partial_{\alpha\dot\beta}\Phi^{(0,1)}{}^{\dot\beta}+\bar M\Phi^{(1,0)}_{\alpha}= 0\,.
\end{equation}
We note however that phase $e^{i\varphi}$ of $M=|M|e^{i\varphi}$ can be absorbed into space-time spinor fields
by the redefinition
$(\Phi^{(1,0)}_{\alpha},\Phi^{(0,1)}{}^{\dot\beta})\rightarrow (e^{i\varphi/2}\Phi^{(1,0)}_{\alpha},e^{-i\varphi/2}\Phi^{(0,1)}{}^{\dot\beta})$.
Thus, the fields (\ref{Ws12-tr}) provide four-component Dirac spinor $(\Phi^{(1,0)}_{\alpha},\Phi^{(0,1)}{}^{\dot\beta})$
providing standard Dirac equation with real mass $m$ and describe spin $1/2$ massive particle.
Finally it can be shown that even for complex mass $M$ the equations (\ref{Dir-eq}) imply Klein-Gordon equations
\begin{equation}\label{KG-Dir-eq}
(\partial^\mu\partial_\mu+m^2)\,\Phi^{(1,0)}_{\alpha}= 0\,,\qquad
(\partial^\mu\partial_\mu+m^2)\,\Phi^{(0,1)}{}^{\dot\beta}= 0\,.
\end{equation}

{\bf Spin 1:}
As the result of twistor transform (\ref{st-fields}) we obtain the following three space-time fields
\begin{equation}\label{Ws-1}
\begin{array}{rcl}
\Phi^{(2,0)}_{\alpha_1\alpha_{2}}(x)&=& {\displaystyle\int d\mu_6(\lambda,\bar\lambda)\,  e^{ix^{\dot\gamma\gamma} \lambda^k_{\gamma}\bar\lambda_{\dot\gamma k}}
\lambda^{i_1}_{\alpha_1}\lambda^{i_{2}}_{\alpha_{2}}\Psi^{(1)}_{i_1 i_{2}}(\lambda,\bar\lambda)}\,,\\[6pt]
\Phi^{(1,1)}{}_{\alpha}^{\dot\beta}(x)&=&{\displaystyle
\int d\mu_6(\lambda,\bar\lambda)\, e^{ix^{\dot\gamma\gamma} \lambda^k_{\gamma}\bar\lambda_{\dot\gamma k}}
\lambda^{i_1}_{\alpha}\bar\lambda^{\dot\beta i_{2}}\Psi^{(1)}_{i_1 i_{2}}(\lambda,\bar\lambda)}\,,\\[6pt]
\Phi^{(0,2)}{}^{\dot\beta_1\dot\beta_{2}}(x)&=&{\displaystyle
\int d\mu_6(\lambda,\bar\lambda)\, e^{ix^{\dot\gamma\gamma} \lambda^k_{\gamma}\bar\lambda_{\dot\gamma k}}
\bar\lambda^{\dot\beta_1 i_{1}}\bar\lambda^{\dot\beta_{2} i_{2}}\Psi^{(1)}_{i_1 i_{2}}(\lambda,\bar\lambda)}\,.
\end{array}
\end{equation}
{}From these definition it follows that these fields satisfy Dirac-like equations
\begin{equation}\label{Dir-eq1a}
i\partial^{\dot\beta\gamma}\Phi^{(2,0)}_{\gamma\alpha}+M\Phi^{(1,1)}{}_{\alpha}^{\dot\beta}= 0\,,\qquad
i\partial_{\alpha\dot\gamma}\Phi^{(0,2)}{}^{\dot\gamma\dot\beta}+\bar M\Phi^{(1,1)}{}_{\alpha}^{\dot\beta}= 0\,,
\end{equation}
\begin{equation}\label{Dir-eq1b}
i\partial^{\dot\alpha\gamma}\Phi^{(1,1)}{}_{\gamma}^{\dot\beta}+M\Phi^{(0,2)}{}^{\dot\alpha\dot\beta}= 0\,,\qquad
i\partial_{\alpha\dot\gamma}\Phi^{(1,1)}{}_{\beta}^{\dot\gamma}+\bar M\Phi^{(2,0)}_{\alpha\beta}= 0\,.
\end{equation}
Further, the formulae (\ref{Dir-eq1a}), (\ref{Dir-eq1b}) even for complex $M$ lead to the Klein-Gordon equations for all fields  (\ref{Ws-1})
\begin{equation}\label{KG-Dir-eq1}
(\partial^\mu\partial_\mu+m^2)\,\Phi^{(2,0)}_{\alpha\beta}= 0\,,\qquad
(\partial^\mu\partial_\mu+m^2)\,\Phi^{(1,1)}_{\alpha\dot\beta}= 0\,,\qquad
(\partial^\mu\partial_\mu+m^2)\,\Phi^{(0,2)}_{\dot\alpha\dot\beta}= 0\,.
\end{equation}
The equations  (\ref{Dir-eq1b}) imply transversality of four-vector field 
$\Phi^{(1,1)}_{\alpha\dot\beta}={\textstyle\frac{1}{\sqrt{2}}}\,\sigma^\mu_{\alpha\dot\beta}A_\mu$
\begin{equation}\label{trans-1}
\partial^{\dot\alpha\beta}\Phi^{(1,1)}_{\beta\dot\alpha}= 0\qquad\leftrightarrow\qquad\partial^\mu A_\mu= 0\,.
\end{equation}
We can consider vector field $\Phi^{(1,1)}_{\alpha\dot\beta}$
as primary field with spin $1$ and remaining two fields $\Phi^{(2,0)}_{\alpha\beta}$, $\Phi^{(0,2)}{}^{\dot\alpha\dot\beta}$ as
derivable from $\Phi^{(1,1)}_{\alpha\dot\beta}$ by the formulae (\ref{Dir-eq1b}) defining selfdual
and anti-selfdual $J=1$ field strengths.
The masses in the equations (\ref{Dir-eq1a}), (\ref{Dir-eq1b}) can be made real after the redefinition
$(\Phi^{(2,0)}_{\alpha\beta},\Phi^{(0,2)}_{\dot\alpha\dot\beta},\Phi^{(1,1)}_{\alpha\dot\beta})\rightarrow
(e^{i\varphi}\Phi^{(2,0)}_{\alpha\beta},e^{-i\varphi}\Phi^{(0,2)}{}^{\dot\alpha\dot\beta},\Phi^{(1,1)}_{\alpha\dot\beta})$,
where $e^{i\varphi}$ is the phase of complex mass $M$.
If we define the $J=1$ field strength (see also \cite{FFLM})
\begin{equation}\label{Proca-fields}
F_{\mu\nu}={\textstyle\frac{im}{\sqrt{2}}}\left(\sigma_{\mu\nu}^{\alpha\beta}\Phi^{(2,0)}_{\alpha\beta}+
\tilde\sigma_{\mu\nu}^{\dot\alpha\dot\beta}\Phi^{(0,2)}_{\dot\alpha\dot\beta}\right)\,,
\end{equation}
due to the equations (\ref{Dir-eq1a}), (\ref{Dir-eq1b}), (\ref{trans-1}) the fields (\ref{Proca-fields}) satisfy
the Proca equations
\begin{equation}\label{Proca-eq}
\partial^\mu A_\mu= 0\,,\qquad \partial_\mu A_\nu-\partial_\nu A_\mu=F_{\mu\nu}\,,\qquad
\partial^\mu F_{\mu\nu}+m^2 A_\nu=0
\end{equation}
and the Bianchi identity $\partial_{[\mu} F_{\nu\rho]}=0$.

For arbitrary $J$ one can derive in analogous way the general form of the Bargmann-Wigner equations for massive fields with arbitrary spin $J$.

\setcounter{equation}{0}
\section{Outlook}

\quad\, Twistor theory aims at providing a new geometric framework for the description of classical and
quantum dynamical models, and one of its basic aims is to formulate the twistor theory of free and
interacting particles. The theory in single $D=4$ twistor space describes conformal 
space-time geometry and provides six-dimensional phase
space of massless particles with remaining two degrees of freedom describing $U(1)$ gauge and discrete set of helicities.
After quantization the twistor theory via so-called twistor transform  provides new method for solving
the field equations for massless fields with arbitrary helicity (see e.g. \cite{EPW}). These techniques were further extended to
curved twistor theory and provided new way of solving Einstein and Yang-Mills equations for selfdual and anti-selfdual cases
(see e.g. \cite{Pen76,Ward}).

The subject studied in this paper is the twistor description of free massive particles with
arbitrary spin. In order to introduce in twistor theory time-like fourmomentum vector it is necessary
to consider the two-twistor geometry, with sixteen real degrees of freedom.
Relativistic spin is described by the Pauli-Lubanski fourvector which carries for definite mass and spin two
new degrees of freedom. These new degrees we describe as parametrizing
two-dimensional fuzzy sphere $\mathbb{S}^2$ with nonAbelian $SU(2)$ Poisson brackets. In this paper we did show that
\begin{itemize}
\item in space-time framework the particle dynamics with nonvanishing spin is obtained adding Souriau-Wess-Zumino term;
\item in order to get pure twistorial formulation of massive particles with spin
we should modify the standard Penrose incidence relations,
which can be obtained by suitable shift of the twistor components;
\item in two-twistor space the model is described by free two-twistor Lagrangian with suitable chosen six constraints bilinear
in twistor variables;
\item the degrees of freedom described by the three-vector $s^r$ due to the constraints (\ref{Sr-s})
can be treated as specifying the choice of conformal-invariant scalar products of the pair of twistors,
i.e. in such a way in physical phase space the variables $s^r$ are determined as well by the twistor components;
\item in order to obtain Pauli-Lubanski spin fourvector one should multiply (see (\ref{W-tw}) and (\ref{S-dot})) the three-vector $s^r$
with internal three-vector indices by three fourvectors $u_\mu^r$ describing the soldering between internal and space-time descriptions of
spin degrees of freedom.
\end{itemize}

The methods presented in this paper can be extended in several ways, in particular to particle models
which generalize the presented here $D\,{=}\,4$ case. In particular
\begin{itemize}
\item one can consider the theory of supersymmetric particles and study the supertwistor description
\cite{Ferb} of free massive superparticles with
nonvanishing superspin. The superspin should be described by supersymmetric extension of
Pauli-Lubanski fourvector \cite{Sok}. The formalism after using the first quantization
rules will provide various known $D\,{=}\,4$ free massive superfields;
\item it should be recalled that infinite higher spin multiplets have been obtained by spinorial
and twistorial formulations of the free particle models in extended space-time with tensorial coordinates 
generated by tensorial central charges
(for $D\,{=}\,4$ the extended tensorial space-time is ten-dimensional \cite{BaLu}--\cite{Vas}).
%\cite{BaLu,BaLuSo,Vas}).
These models used only the set of single twistorial variables and were describing massless higher spin fields.
It is interesting to consider the massive two-twistor models linked with tensorially extended space-time
which can be obtained by dimensional reduction of higher-dimensional massless
spinorial theory in extended tensorial space-time. This idea has been already outlined in our previous paper \cite{FedLuk}
(see also \cite{SV}), with the description
of two-twistor $D\,{=}\,3$ massive spinorial model as obtained by the dimensional reduction from
$D=4$ massless spinorial model.
\end{itemize}

\section*{Acknowledgements}

\noindent We acknowledge a support from the
grant of the Bogoliubov-Infeld Programme and RFBR grants 12-02-00517, 13-02-91330 (S.F.), as
well as Polish National Center of Science (NCN) research projects
No.~2011/01/ST2/03354 and No.~2013/09/B/ST2/02205 (J.L.). S.F. thanks the members of the Institute of Theoretical Physics
at Wroc{\l}aw University for their warm hospitality.

\end{document}